\documentclass[final,5p,times,twocolumn]{elsarticle}

\usepackage{graphicx}
\graphicspath{{./figures/}}

\usepackage{amssymb}
\usepackage{url}
\usepackage{float}

\ifcsname italic \endcsname%
\else%
  
\fi
  
\journal{Nuclear Instruments and Methods A}

\begin{document}

\begin{frontmatter}



\title{The OLYMPUS Internal Hydrogen Target}


\author[mit]{J.~C.~Bernauer\corref{cor}}
\ead{bernauer@mit.edu}

\author[fer]{V.~Carassiti}

\author[fer]{G.~Ciullo}

\author[mit]{B.\ S.~Henderson}

\author[mit,bates]{E.~Ihloff}

\author[mit,bates]{J.~Kelsey}

\author[fer]{P.~Lenisa}

\author[mit,bates]{R.~Milner}

\author[mit]{A.~Schmidt}

\author[fer]{M.~Statera}

\cortext[cor]{Corresponding author}

\address[mit]{Massachusetts Institute of Technology, Laboratory for Nuclear Science, Cambridge, MA 02139, USA}
\address[bates]{MIT-Bates Linear Accelerator Center, Middleton, MA 01949, USA}
\address[fer]{Istituto Nazionale di Fisica Nucleare and Universit\`{a}, 44100, Ferrara, Italy} 

\begin{abstract}

  An internal hydrogen target system was developed for the OLYMPUS
  experiment at DESY, in Hamburg, Germany.  The target consisted of a
  long, thin-walled, tubular cell within an aluminum scattering
  chamber.  Hydrogen entered at the center of the cell and exited
  through the ends, where it was removed from the
  beamline by a multistage pumping system. A cryogenic coldhead cooled
  the target cell to counteract heating from the beam and
  increase the density of hydrogen in the target. A fixed collimator
  protected the cell from synchrotron radiation and the beam halo. A
  series of wakefield suppressors reduced heating from beam
  wakefields.  The target system was installed within the DORIS
  storage ring and was successfully operated during the course of the
  OLYMPUS experiment in 2012. Information on the design, fabrication,
  and performance of the target system is reported.

\end{abstract}


\end{frontmatter}



\section{Motivation}

The OLYMPUS experiment \cite{Milner:2013daa} at DESY, in Hamburg,
Germany, was designed to measure the 
cross section ratio of electron-proton to positron-proton elastic scattering over the range of negative four-momentum 
transfer from 0.4 to 2.2~$($GeV$/c)^2$ at a fixed beam energy of 2.01~GeV. A contribution to the cross
section from hard two-photon exchange would cause this ratio to deviate from unity. Such contributions are
believed to at least partially resolve the discrepancy in the proton form factor ratio 
$G_E/G_M$ measured in Rosenbluth separation and polarization
experiments \cite{Guichon:2003qm}. There is no theoretical
consensus on the size of the two-photon contribution \cite{Arrington:2011dn},
and a direct experimental measurement is needed. The design goal for OLYMPUS was
to measure this ratio with less than 1\% uncertainty.

\begin{figure}[tbh]
\centering
\includegraphics[width=0.8\columnwidth]{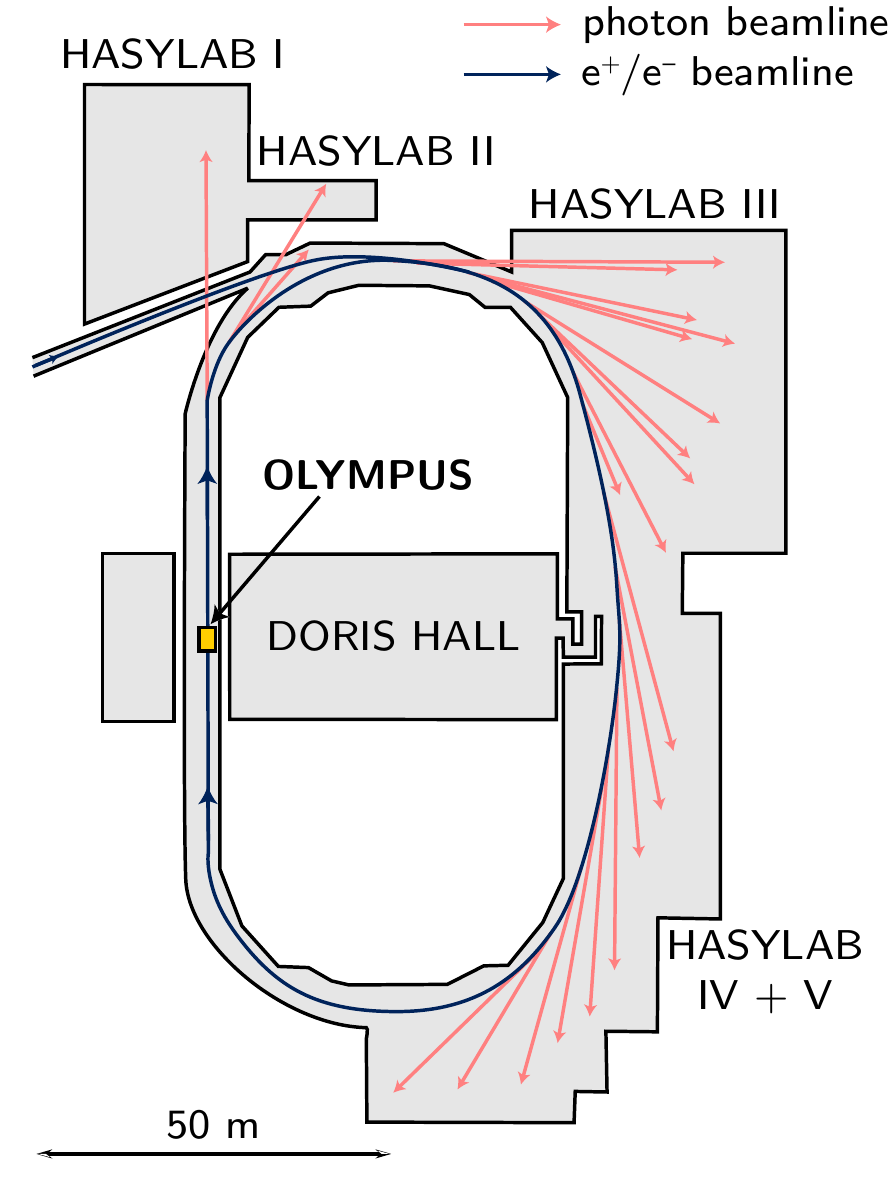}
\caption{The layout of the DORIS $e^+/e^-$ storage ring at DESY, showing the position
of the OLYMPUS detector and the synchrotron light experiment stations.}
\label{fig:doris}
\end{figure}

The OLYMPUS detector was installed at the DORIS storage ring at DESY, the
layout of which is shown in Fig.\ \ref{fig:doris}. 
DORIS was selected because it could circulate both electron and positron beams 
with multi-GeV energies and currents up to 150~mA. OLYMPUS aimed to 
measure the elastic cross section ratio by directing the DORIS beam
through an internal hydrogen gas target, switching beam species about once
per day. The scattered lepton and the recoiling proton were
detected in coincidence. 
For its detector systems, OLYMPUS incorporated several components 
originally used in the Bates Large Acceptance Spectrometer Toroid (BLAST) experiment
\cite{Hasell:2009zza}.  The differences in physics goals and beam
environments between OLYMPUS and BLAST, however, required the design of 
a new unpolarized hydrogen target system. The target system faced three
principal technical challenges:

\begin{enumerate}
\item In order for OLYMPUS to attain a sufficient luminosity, a target thickness of $3\times 10^{15}$~atoms/cm$^2$ was required.
\item The high beam current made it infeasible to have the beam pass through
target windows. Therefore, a multistage pumping system was required to avoid spoiling the DORIS
beamline vacuum with hydrogen from the target.
\item The bunches in DORIS could carry as much as 30~nC each. With such a high 
charge per bunch, any conducting structures close to the beam were susceptible 
to intense wakefield heating. To reduce the heat load, extensive
wakefield suppression was needed.
\end{enumerate}

To meet these technical demands, the OLYMPUS target was designed as
a windowless gaseous target, internal to the DORIS ring. The DORIS beam passed through
an elliptical tubular target cell into which H$_2$ gas was flowed. The gas was fed into the center of the
cell and was removed by a three-stage differential pumping system as it flowed out the ends of the cell.
A fixed tungsten collimator was positioned in front of the cell to protect it from
synchrotron radiation and the beam halo. The cell, collimator, and 
beam pipe were all linked by conducting transitions to suppress wakefields.
The cell was cooled by a cryogenic coldhead to increase the target density
and to remove heat produced by wakefields.

When not providing a beam for OLYMPUS, the DORIS storage ring served as 
a synchrotron light source. During OLYMPUS data taking, DORIS circulated 
40 to 65~mA beams in ten bunches at an energy of 2.01~GeV. In light source mode, 
DORIS circulated 140~mA positron beams in five bunches at an energy of 4.45~GeV. 
Due to the impracticality of removing the target system, the OLYMPUS target was
designed to tolerate the significantly more taxing conditions of light source
operation.

\section{Design}

\subsection{Scattering Chamber}

\begin{figure}[H]
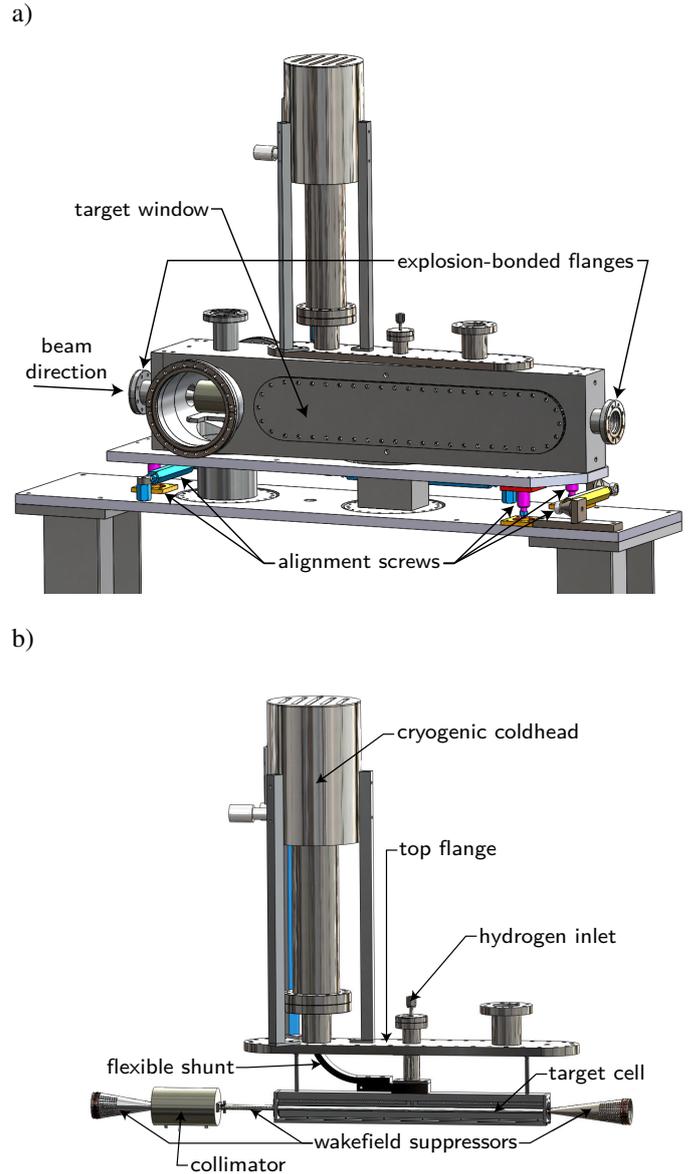

a)
\begin{center}
\includegraphics[width=\columnwidth]{scattering_chamber_labels.pdf}
\end{center}
b)
\begin{center}
\includegraphics[width=\columnwidth]{chamber_inside.pdf}
\caption{\label{fig:chamber}Design drawings of the scattering chamber (a), and the inner target system
components and coldhead (b)}
\end{center}
\end{figure}

The target cell, collimator, and wakefield suppressors were housed
within an aluminum scattering chamber, which is shown in Fig.\
\ref{fig:chamber}a. The DORIS beam was directed through the chamber via
ports on the upstream and downstream faces of the chamber. The
interior of the chamber was directly open to the ring vacuum.  The
chamber was 1.2~m long and had a front face that was 245~mm wide and
254~mm high. It was tapered to a width of 114.3~mm at the downstream
end to increase the visibility of the target cell to the forward
detector elements. The chamber was manufactured from a solid block of
aluminum to ensure high vacuum integrity.

The upstream and downstream beamlines were made of stainless steel pipes. In order to connect the
aluminum of the scattering chamber to the steel of the beam pipes, the upstream and downstream
ports had Atlas\footnote{Atlas Technologies, Port Townsend, WA, USA.} explosion-bonded 
bimetallic flanges. The aluminum sides of these flanges were welded to the scattering chamber. 
The steel sides were clamped to the flanges of the beam pipe, with
copper gaskets in between.

The top of the scattering chamber featured a long port for installing
the target cell. The top flange included feedthroughs for the hydrogen
gas supply, the cryogenic coldhead, and various sensors. The lateral
faces of the scattering chamber had large access ports for installing
the collimator as well as two long windows to allow scattered
particles to exit.  The components inside the scattering chamber are
shown in Fig.\ \ref{fig:chamber}b.

The windows on each side of the chamber were made from  0.25~mm thick
1100 aluminum alloy foil, and   
were sealed using O-rings.  The windows subtended a polar
angle range (measured from the center
of the cell relative to the beam direction) of $8^\circ$ to $100^\circ$, providing
a large acceptance range for the detector. The trapezoidal shape of the chamber was
chosen so that the windows extended to scattering angles below $12^\circ$, giving the
experiment's $12^\circ$ luminosity monitors an unobstructed line-of-sight to most of the target.
The trapezoidal shape also angled the windows so that leptons scattered at forward
angles passed through less window material, reducing energy loss and multiple
scattering.

\subsection{Target Cell}

\begin{figure}[htb]
\centering
\includegraphics[width=\columnwidth]{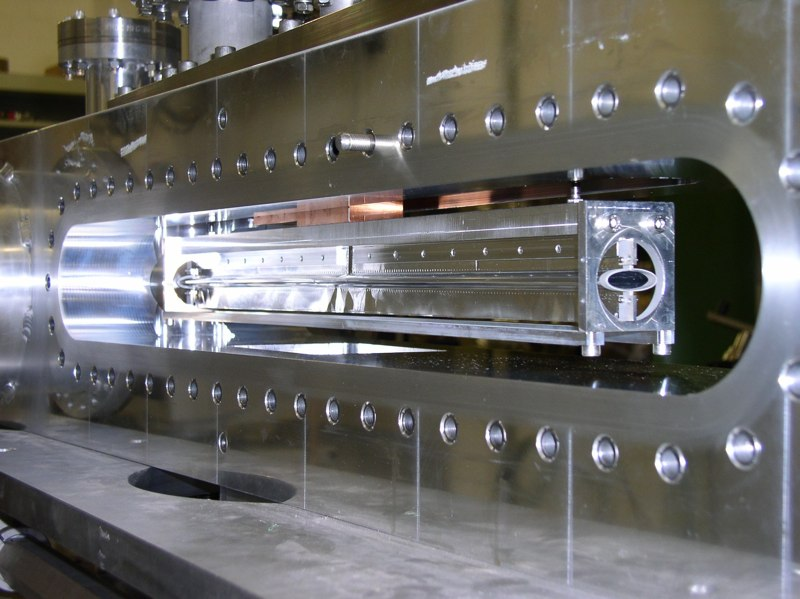}
\caption{The target cell and frame are shown mounted in the scattering chamber, with the 
target windows removed.}
\label{fig:cell}
\end{figure}

The target cell, shown in Fig.\ \ref{fig:cell}, was a 600 mm long
aluminum tube through which the DORIS beam was directed. The cell had
an elliptical cross section (27~mm wide by 9~mm high) to match the
aspect ratio of the beam envelope \cite{Brinker:2009}. The tube walls
were made of thin (75~$\mu$m) aluminum sheets to reduce multiple
scattering of outgoing particles. At the center of the tube, a small
port was molded between the sheets to provide an inlet for the
hydrogen gas.  Once inside the cell, gas could flow only out through
the ends of the tube where it was removed from the beamline by the
vacuum system.

Several target cells were constructed at the Ferrara University/INFN mechanical workshop, 
with the experience gained from the fabrication of cells for the target of the HERMES experiment
\cite{Airapetian:2004yf}. Cells were formed by the following procedure:

\begin{itemize}

\item Two sheets of 75~$\mu$m Goodfellow\footnote{Goodfellow Corporation, Coraopolis, PA, USA} 99.5\% pure Al foil were cut,
cleaned with ethanol, and then thermally softened for two hours at 270~$^{\circ}$C between thermic glass plates to prevent
foil ripples. Each sheet was 600~mm long and 200~mm wide.

\item The sheets were aligned on a copper mold and formed with a pneumatic press. 
The foils are shown before and after the pressing process 
in Fig.\ \ref{fig:constr_fig2}. During the process, the two halves of the gas injection
tube at the center of the cell were also formed.

\begin{figure}[ht]
a)
\begin{center}
\includegraphics[width=\columnwidth, trim=1cm 3.7cm 1cm 3.7cm, clip=true]{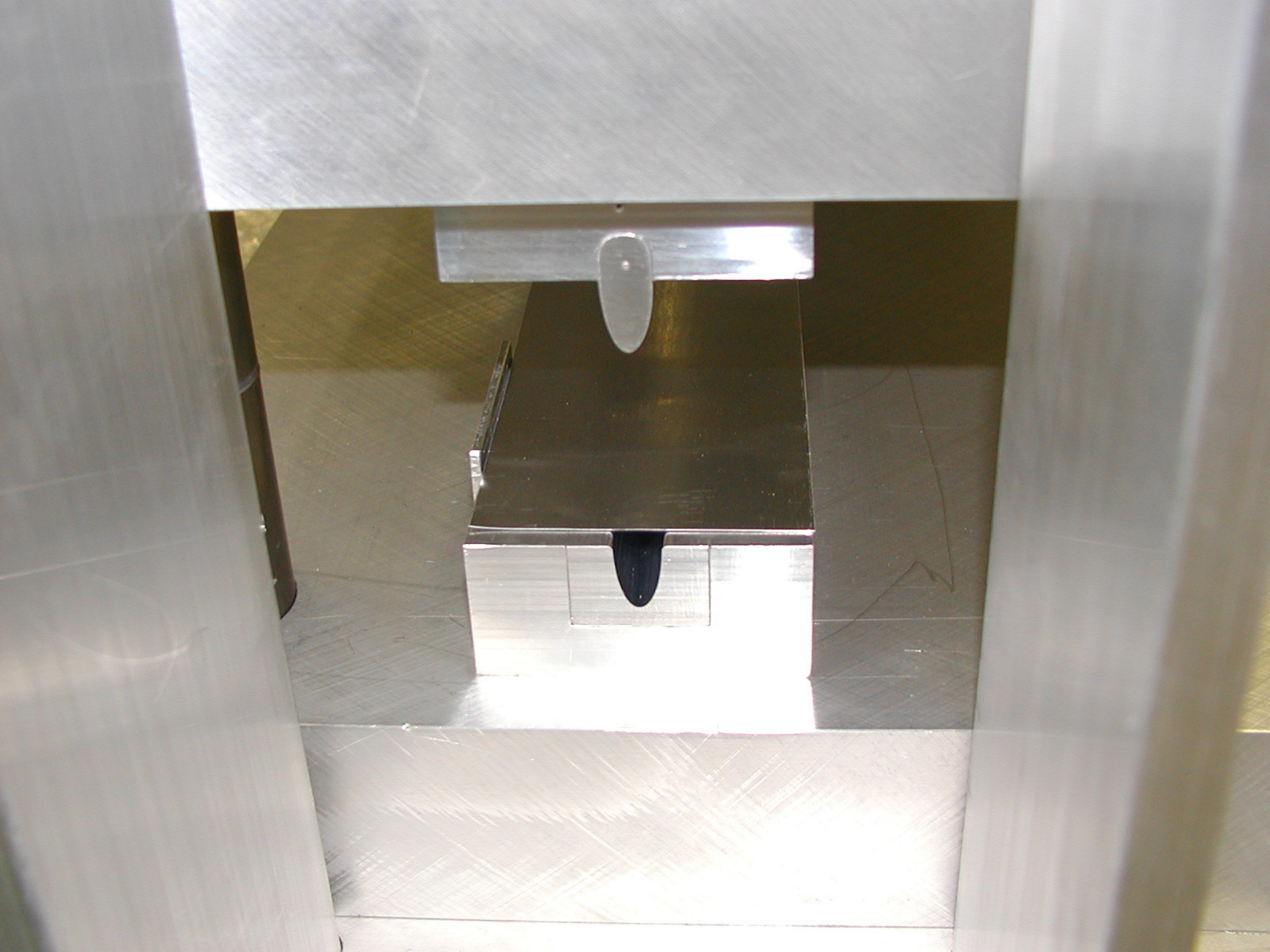}
\end{center}
b)
\begin{center}
\includegraphics[width=\columnwidth, trim=8cm 8cm 8cm 8cm, clip=true]{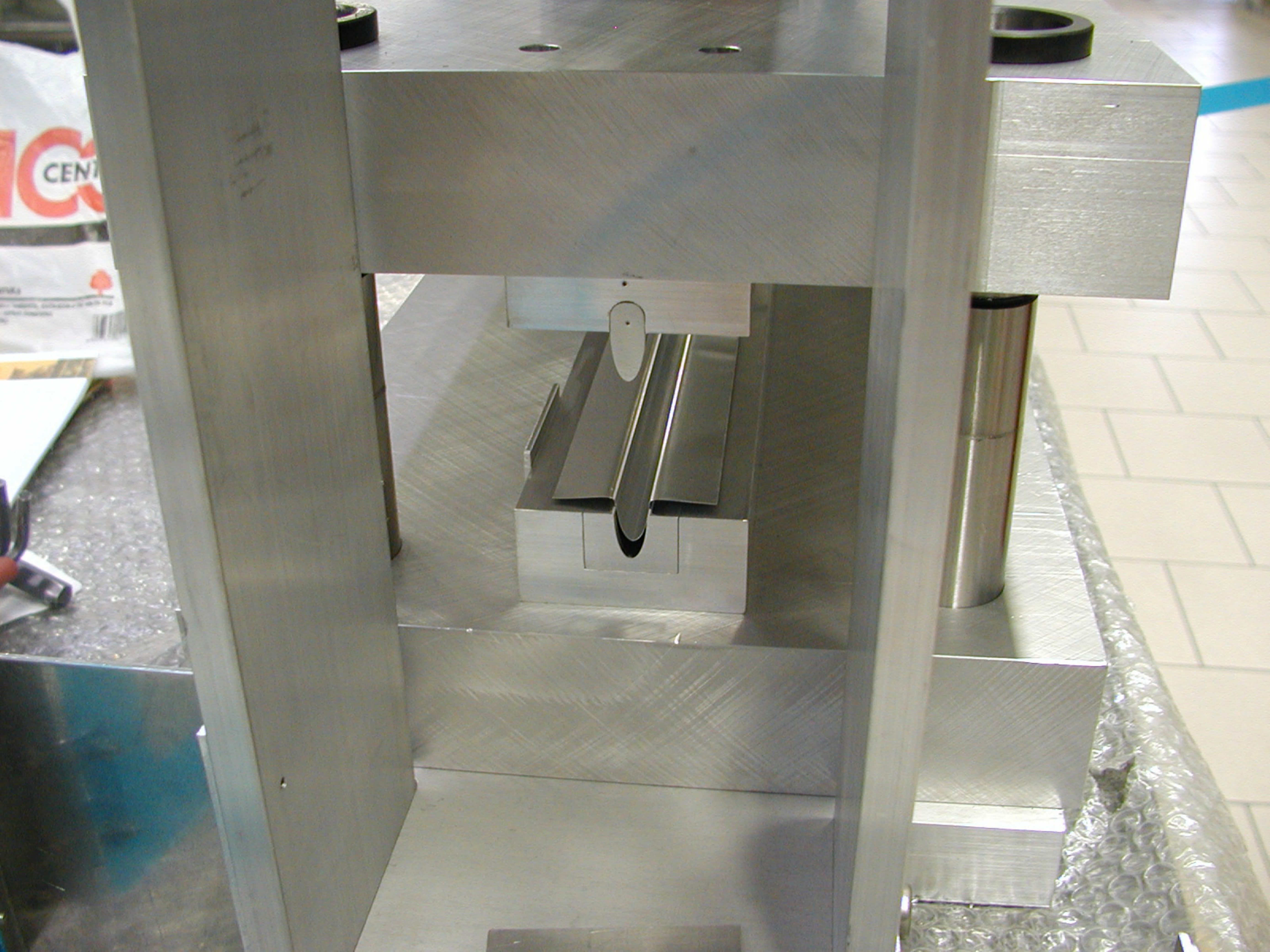}
\caption{The target cell foil in the pneumatic press before (a) and after (b) shaping \label{fig:constr_fig2}}
\end{center}
\end{figure}

\item After molding, the ends of each sheet were fitted with two
  aluminum reinforcement pieces, as shown in Fig.\
  \ref{fig:constr_fig3}.  Additionally, the reinforcements provided
  support for clamping the wakefield suppressors to the cell.

\begin{figure}[ht]
\centering
\includegraphics[width=\columnwidth]{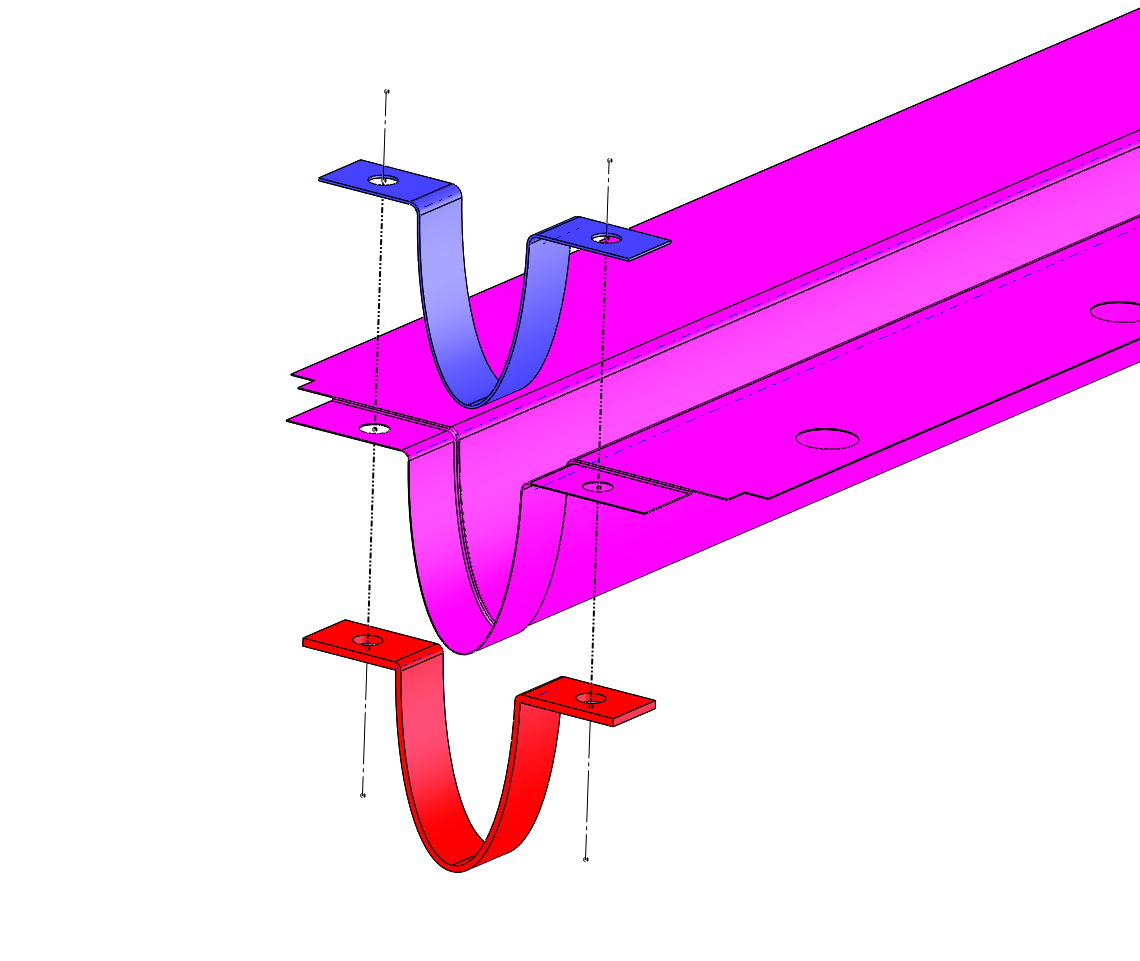}
  \caption{Schematic of the reinforcement pieces at the ends of the cell}
\label{fig:constr_fig3}
\end{figure}

\item The two cell halves were then placed on a copper mold and spot welded together
along the top and bottom seams. The mold served as one of the electrodes for the
spot welding.

\end{itemize}

After a cell was fabricated, it was fixed into a mounting
assembly. First, the cell was secured in an aluminum frame made of the high-purity 6063 aluminum alloy, 
which has excellent thermal conductivity at cryogenic temperatures. The frame 
provided the rigid support for a wrapping of ten layers of aluminized Mylar 
reflective insulation. A copper block, which served as a high-conductivity thermal junction, 
was attached to the top of the frame at the midpoint of the cell. The 
cell assembly was suspended from the top flange by a steel 
tube. Pins at each end of the top flange prevented the
ends of the cell assembly from bending upward while under thermal stress.

The cell's position in the scattering chamber was adjusted by adding shims 
into the mounting assembly. Final adjustments of the cell's position were made by 
turning alignment screws on the scattering chamber, which moved the entire chamber 
relative to the beam. 

\subsection{Collimator}

A cylindrical tungsten collimator, placed directly in front
of the target cell, shielded the cell
from synchrotron radiation and from the beam halo. It was
fabricated from a solid cylinder of tungsten using wire 
electrical discharge machining. The collimator had the smallest
aperture in the DORIS ring. The opening in the upstream face
was elliptical, 25~mm wide by 7~mm high (approximately
the $10\sigma$ beam diameter at that location for a 4.5~GeV 
beam \cite{Brinker:2009}). The aperture flared towards the downstream
end, to 27~mm wide by 9~mm high, to prevent small-angle scattering of
synchrotron photons off the interior surface of the collimator onto the cell walls.

\begin{figure}[htb]
\centering
\includegraphics[width=\columnwidth]{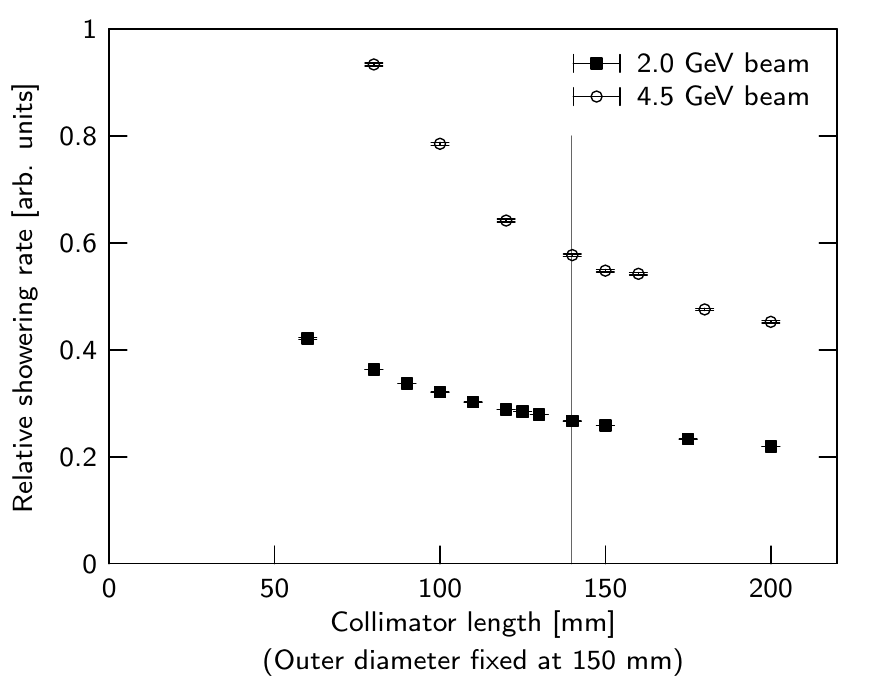}\\
\includegraphics[width=\columnwidth]{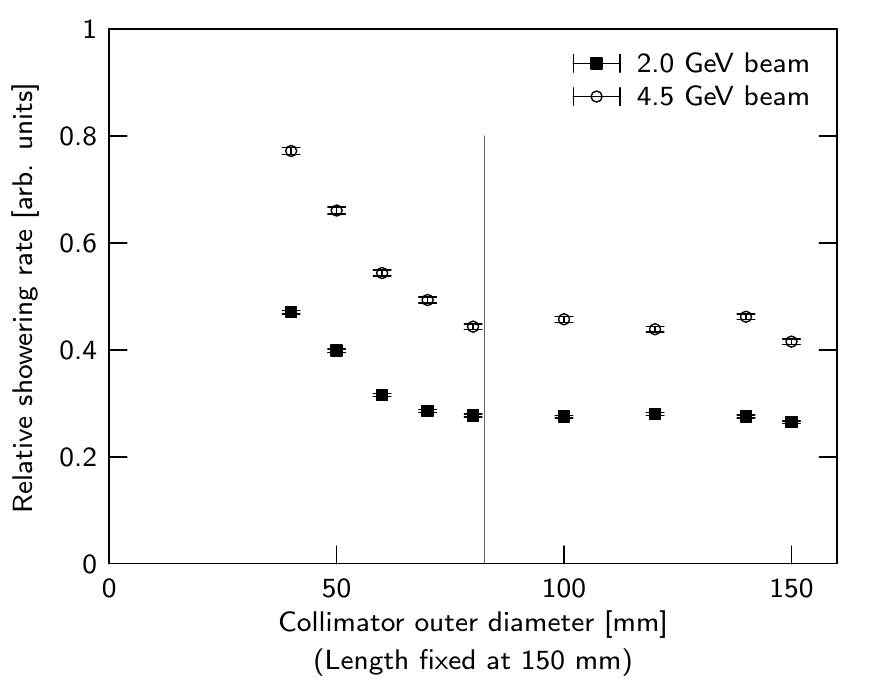}
\caption{A Monte Carlo simulation was used to study how the dimensions of the collimator
affected the showering rate from beam halo particles. A length of 139.7~mm and an 
outer diameter of 82.55~mm were chosen for the final design.}
\label{fig:coll}
\end{figure}

The outer dimensions of the collimator were selected after taking into 
account the results of a  Monte Carlo simulation, in which 
electrons and positrons from the beam halo were made to impinge against 
the upstream face of the collimator. The particles from the resulting 
shower that managed to exit the collimator and the scattering chamber were 
counted. After varying the collimator's dimensions in the simulation,
an outer diameter of 82.55~mm and a length of 139.7~mm were selected, 
balancing weight and cost with shielding performance. The results of the 
simulation are shown in Fig.\ \ref{fig:coll}.

The heat load on the collimator due to incident synchrotron
radiation was estimated to be on the order of 25~W  during 4.5~GeV
running, which would easily be dissipated through
the collimator supports. During 2~GeV running, the power would be 
on the order of 1~W.

\subsection{Wakefield Suppressors}
\label{sec:wakefield}

Wakefield suppressors were necessary to maintain the target cell at cryogenic
temperatures by preventing heating caused by beam wakefields. The wakefield suppressors 
spanned gaps between conducting structures surrounding the beam to
ensure continuous conductivity.  Any sharp transitions
or gaps in conductivity would act as electromagnetic cavities that would be
excited by the passing beam, causing heating
in the surrounding elements. To
prevent this, three wakefield suppressors were produced to cover the following transitions:
\begin{enumerate}
  \item From the upstream 60~mm diameter circular scattering chamber port to the
    25~mm by 7~mm elliptical opening of the collimator,
  \item From the exit of the collimator to the entrance of the target
    cell (both 27~mm by 9~mm elliptical), and
  \item From the 27~mm by 9~mm elliptical exit of the target cell to
    the 60~mm circular diameter of the downstream scattering chamber port.
\end{enumerate}
The design goal was to limit wakefield heating to 10~W.

The upstream wakefield suppressor was screwed to the upstream face of the 
collimator.
A beryllium-copper (BeCu) spring cone made a sliding connection with
the inside of the upstream chamber port to provide good electrical
contact despite thermal expansion and contraction. Since the 
wakefield suppressor was unshielded by the collimator and
needed to be tolerant to synchrotron radiation and beam halo particles, it was
fabricated from aluminum. Aluminum's high thermal conductivity helped 
to dissipate heat from radiation, while its low density offered a
comparatively small
cross section for beam halo particles. The wakefield suppressor had
a thin coating of silver to increase its electrical conductivity. 

\begin{figure}[htb]
\centering
\includegraphics[width=\columnwidth]{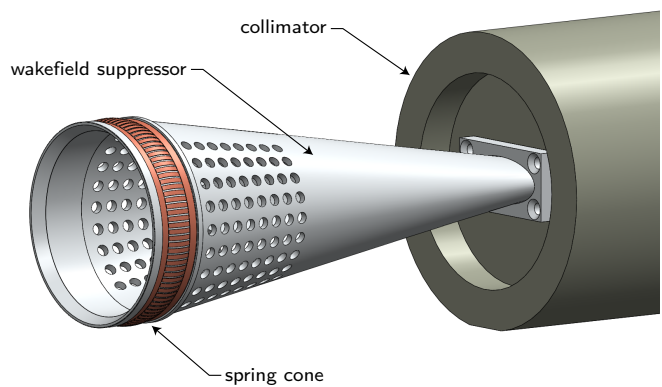}
\caption{Design drawing of the upstream wakefield suppressor, which bridged the
gap between the upstream beam pipe and the collimator}
\label{fig:wakefield_up}
\end{figure}

\begin{figure}[htb]
\centering
\includegraphics[width=\columnwidth]{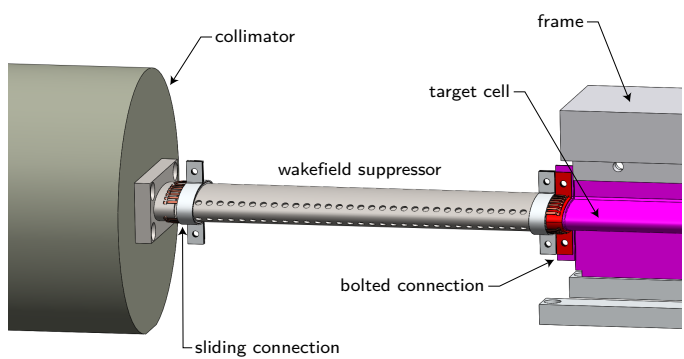}
\caption{Design drawing of the middle wakefield suppressor, which bridged the
gap between the collimator and the target cell}
\label{fig:wakefield_mid}
\end{figure}

\begin{figure}[htb]
\centering
\includegraphics[width=\columnwidth]{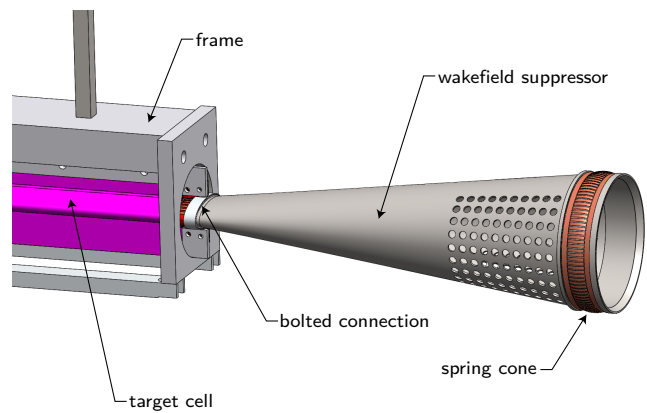}
\caption{Design drawing of the downstream wakefield suppressor, which bridged the
gap between the target cell and the downstream beam pipe}
\label{fig:wakefield_down}
\end{figure}

The middle and downstream wakefield suppressors, shown in Figs. \ref{fig:wakefield_mid}
and \ref{fig:wakefield_down}, were made from silver-coated stainless steel. The middle wakefield
suppressor was fixed with a bolted connection to the reinforcement piece at the upstream end of 
the target cell. BeCu spring fingers extended 
across the connection to maintain good electrical conductivity, even
when thermal contraction produced a gap. The upstream end also had BeCu spring fingers, which made a sliding
connection to an elliptical spout that was screwed to the downstream
face of the collimator. The downstream wakefield suppressor was fixed with a bolted connection to the
reinforcement piece at the downstream end of the target cell. BeCu
spring fingers were used to maintain electrical conductivity across
the connection. The wakefield suppressor's downstream end had a BeCu
spring cone, which made a sliding connection to the downstream chamber
port.

The wakefield suppressors were machined to include numerous small
holes through which the target gas could escape into the scattering chamber.  The
locations of the holes were chosen to be as far as possible from the
beam to reduce wakefield heating from the holes.

\subsection{Cryogenic System}

A cryogenic system was used to remove heat from the cell and
to increased the target's density by reducing the cell's gas conductance. Since the hydrogen
gas entered the cell in the molecular flow regime, it reached equilibrium temperature with
the cell within a few collisions with the cell walls. Geometric calculations of the conductance
of the cell suggested that a target temperature of 75 K would be needed to attain a
target thickness of $3\times 10^{15}$ atoms/cm$^2$ at a flow rate of 0.6~sccm into the cell.

The cell was cooled by a CryoMech\footnote{CryoMech, Inc., Syracuse, NY, USA}  AL230 coldhead with a CP950 compressor. 
The coldhead was mounted directly above the scattering chamber so that it
could be close to the cell but out of the way of the OLYMPUS detectors positioned
in the horizontal plane. In this position, the coldhead was in a region of high magnetic
field, so a cylindrical steel covering was placed over the coldhead to provide
magnetic shielding. The compressor was located in the pit below the
experiment and did not require magnetic shielding.

The coldhead's heat exchanger was coupled to the copper block on the
cell assembly 
by a flexible laminated copper shunt, manufactured by Watteredge\footnote{Watteredge, Inc., Avon Lake, OH, USA}. 
The shunt's flexibility accommodated the expansion and contraction of
the target during changes in temperature and dampened mechanical vibration from the coldhead. A thin 
layer of indium was placed between all of the thermal couplings to improve heat conductivity. 

The scattering chamber remained at room temperature and needed to be thermally insulated 
from the cell. The cell assembly was suspended from the chamber by a steel tube, which 
was fixed to the copper block by four small screws. The tube was thin, to reduce its 
thermal conductivity. Sheets of Kapton were placed between the tube and the copper block 
to provide insulation. 

\begin{figure}[htb]
\centering
\includegraphics[width=\columnwidth]{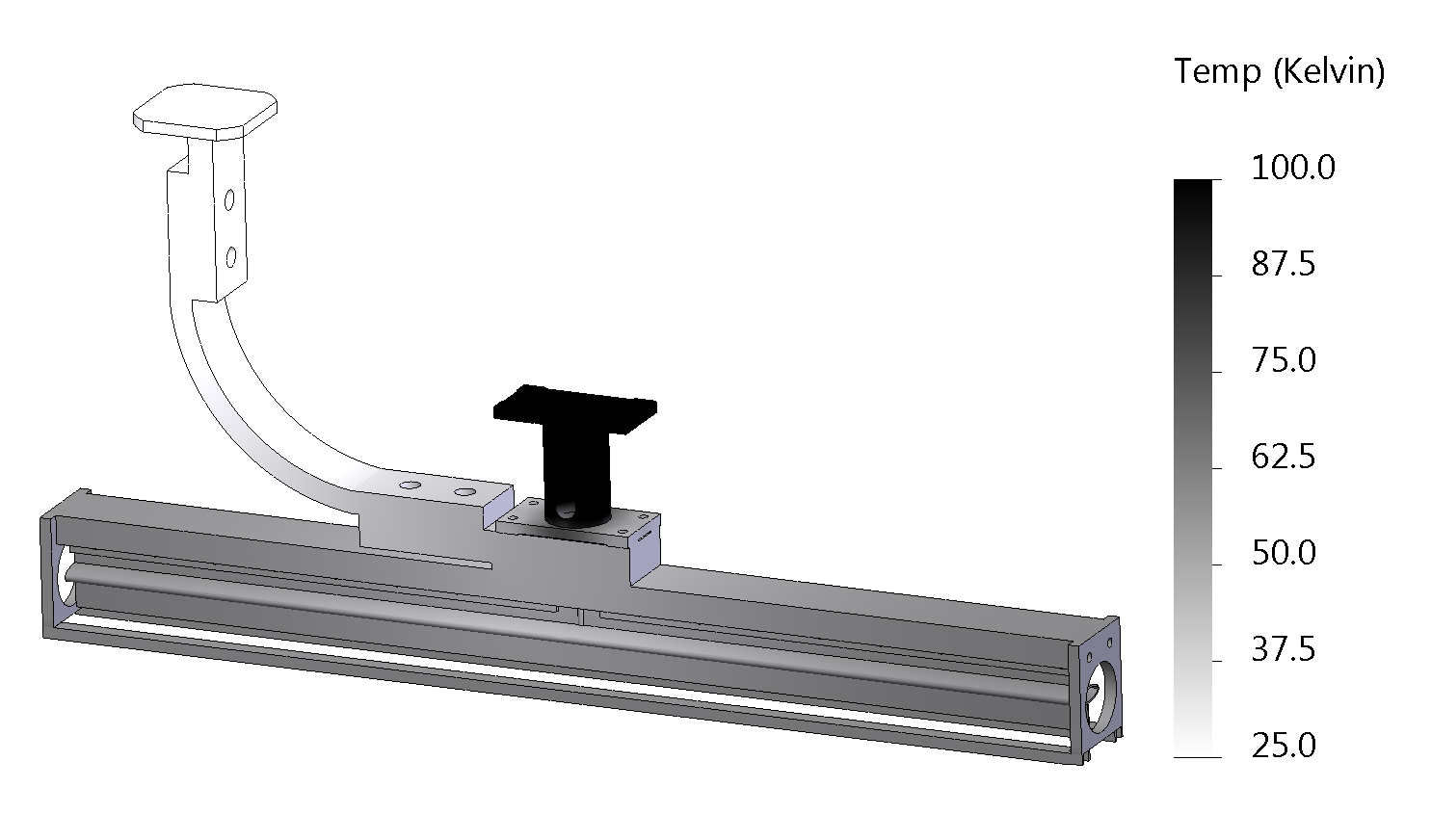}
\caption{The temperature profile of the target cell was simulated
  assuming a heat load of 10~W on the cell.  The simulation
suggested that the coldhead could maintain a target temperature of 75~K.}
\label{fig:heat}
\end{figure}

The coldhead was rated to dissipate 36~W of thermal power at a temperature 
of 25~K. A simulation of the thermal conductance of the target components,
assuming 10~W of thermal power deposited on the cell, indicated that a target temperature of 75~K
could be sustained. The results of this calculation are shown in Fig.\ \ref{fig:heat}.

The temperature of the target cell was monitored by seven Pt100 temperature sensors.
The sensors were positioned along the length of the cell so that the temperature 
uniformity could be monitored. The wires for the sensors were fed through the ports
in the top of the scattering chamber.

\subsection{Vacuum System}

\begin{figure}[htb]
\centering
\includegraphics[width=\columnwidth]{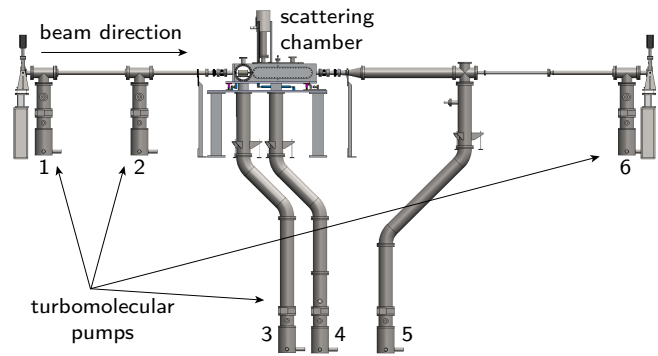}
\caption{The vacuum system as oriented in the DORIS beamline.}
\label{fig:vacuum}
\end{figure}

A system of vacuum pumps removed the hydrogen from the scattering chamber and beamline
after it exited the target cell. Since the cell was internal to the beamline,
the vacuum system was critical to the operation of the the target as it prevented the hydrogen from the target
from spoiling the vacuum of the DORIS ring. The system included six turbomolecular
vacuum pumps (Osaka\footnote{Osaka Vacuum Ltd., Osaka, Japan.} TG 1100M and 
Edwards\footnote{Edwards, Crawley, UK.} STP 1003C), each with 800~L/s capacity for hydrogen, and four
Non-Evaporable Getters (NEGs) (SAES\footnote{SAES Group, Lainate, Italy.}
CapaciTorr CFF 4H0402), each with 400~L/s equivalent capacity for hydrogen. Fig.\ \ref{fig:vacuum} 
shows the placement of the pumps in relation to the beamline.

The vacuum system design was based on a detailed conductance map of all the
vacuum components around the OLYMPUS section of the beamline. The pumps were arranged
in three stages, with each stage reducing the pressure by an order of magnitude, from 
$10^{-6}$~Torr in the scattering chamber to $10^{-9}$~Torr in the DORIS ring. The first
stage pumps (3 and 4), were positioned below the scattering chamber, the second stage 
pumps (2 and 5) were positioned upstream and downstream of the scattering chamber, while the 
third stage pumps (1 and 6) were positioned at the beginning and end of the OLYMPUS 
section of the beamline. NEGs were placed above the outer four pumps (1, 2, 5, and 6) 
to assist in removing hydrogen at low pressure. 

Since all of the turbomolecular pumps had magnetically levitating rotors, they could not
operate in a magnetic field. Pumps 3, 4, and 5 were placed in the pit, several meters below 
the beamline, outside the fringe field of the OLYMPUS toroid magnet. 
To compensate for the additional pumping distance, 20~cm diameter pipes were 
used to connect the pumps to the beamline with sufficient conductance.

\subsection{Gas Feed System and Supply}

The target system employed a Parker\footnote{Parker-Hannifin
  Corporation, Haverhill, MA, USA} 75-34 hydrogen gas generator, which
produced ultra-pure molecular hydrogen (fewer than 200 ppb of impurities) via
electrolytic dissociation of water.  To create the ultra-pure
hydrogen, the generator incorporated a cathode consisting of palladium
tubes, into which only H$^+$ ions could pass. The ions combined to
form H$_2$ in these tubes before exiting the generator to the supply
line. Only water deionized to a resistivity of greater than 5
M$\Omega$-cm was used in the system to ensure purity and to protect 
the palladium cathode.  The generator
operated continuously throughout OLYMPUS data taking to maintain a
pressure of 20~psi on the supply line, so as to keep the line at
positive pressure relative to the atmosphere and to provide an ample
supply to the gas feed system.

The flow of gas into the target cell was controlled by a
remotely-operated system of mass flow controllers (MFCs) and pneumatic
solenoid valves, modeled on the system used for the BLAST experiment's
target \cite{Cheever:2006xt} at the MIT Bates Linear Accelerator.  The
MFCs provided reliable flow control in the range from approximately
0.1 to 1~sccm.  Two ``buffer/reservoir'' tanks of precisely-known
volume were attached to the system to provide a means of calibrating
the MFC output.  Additionally, the volumes could be used to drive the
output flow using the differential pressure between the volumes, which
allowed more precise control of lower flow rates.  The valves and MFCs
were controlled during experimental running via the slow control
system described in Section \ref{slowctrl}, but were also controllable
using a console system beneath the OLYMPUS detector.

\subsection{Slow Control System}
\label{slowctrl}
The target slow control system used VME-based CPUs and
modules. Standard VME I/O modules were used to control the valves of
the gas feed system (using custom made valve control boxes) and to read
the status of the turbo pump controllers.  VME ADC cards were used to
read the output of the pressure sensors in the target supply system
and the flows measured by the MFCs. The MFC flow was set by VME DAC
cards. The pressure sensor modules of the target system were equipped
with RS-232 serial interfaces, which were connected to the VME CPU via
off-the-shelf RS-232 to USB converters.

The software of the slow control system was divided into three parts:
the interface to the hardware, a database, and the user interface. The
interface to hardware was realized using the Experimental Physics and
Industrial Control System (EPICS)\footnote{Experimental Physics and
  Industrial Control System, Argonne National Laboratory, USA,
  \url{http://www.aps.anl.gov/epics/index.php}} software package,
which also maintained an image of the current system state. The system
state was replicated in a SQL database, which also recorded the
history of the various data channels. The database backend was
realized using PostgreSQL. The main DAQ system of OLYMPUS queried the
database at constant time intervals and updated the experiment data
files with information from the slow control channels.  Custom
interconnection daemons connected the database to EPICS and the slow
control system of DESY, called TINE\footnote{Three-fold Integrated
  Networking Environment, DESY, Germany, \url{http://tine.desy.de}}.
The custom graphical user interface was built on a web-based
platform. A server, written in Python using the FLASK framework,
provided an HTML- and JavaScript-based GUI, which was accessible from
any computer with a web browser inside the DESY network. View-only
access from outside DESY was realized with a replicated database and a
read-only version of the server running on an exposed host, a
virtual machine provided by the DESY IT department.

The complete slow control design, originally developed for the target
system alone, was later adopted and extended to control all aspects of
the OLYMPUS experiment (see \cite{Milner:2013daa}).

\section{Operations}

A prototype of the OLYMPUS target system was installed in the DORIS ring in
January 2011 for a test experiment. At the end of this
test, the prototype target was uninstalled and improvements were made based on
the experience gained. The full target system was installed
in July 2011, along with the OLYMPUS spectrometer. Commissioning tests took
place during the fall of 2011. The first OLYMPUS data run took place during
February 2012. The second OLYMPUS data run took place from October 2012 to
the beginning of January 2013. 

\subsection{Test Experiment}

A test experiment was conducted with a prototype target system in
January and February 2011. Since the OLYMPUS detector was not yet
assembled, some simple detectors were arranged to detect leptons and
protons in coincidence from elastic scattering in the target: a
telescope of multi-wire proportional chambers and a calorimeter of lead-glass
crystals were used to detect scattered leptons at $12^\circ$, and an
array of three time-of-flight scintillator bars detected recoiling
protons at back angles. During these tests, the cryogenic and vacuum
systems performed well under OLYMPUS-like running
conditions. Reconstruction of the scattering vertices showed that the
target had the expected triangular density profile.

\subsection{Synchrotron Operation}

After the test experiment, DORIS resumed synchrotron operation. The
prototype target was not able to withstand the conditions of
synchrotron operation, became damaged, and had to be
removed. Several flaws were identified and corrected in the final
design, which was described in Section \ref{sec:wakefield}.  The
prototype target cell frame, which was not built from 6063 aluminum,
began to warp under thermal stress. The prototype wakefield
suppressors had gaps of approximately 1~mm at each junction, which
caused excessive heating. These gaps were eliminated in the final design by
the system of bolted and sliding connections. The laminated copper
shunt replaced a simple copper braid to improve the heat
conductivity between the target cell frame and the coldhead.

The improved target system was installed in July 2011. It performed well during
OLYMPUS commissioning tests and withstood synchrotron operation, during which
the target cell temperature was approximately 50~K. For comparison, the target 
achieved a temperature of approximately 40~K with no beam.

\subsection{OLYMPUS Run I}

During the first OLYMPUS run, data were collected with beam currents between
40 and 65~mA, divided in ten bunches. The target was operated with a flow of 0.8~sccm
while maintaining a satisfactory beam lifetime on the order of 30~minutes.
In these conditions, the target had a temperature of approximately 55~K.

Analysis of the luminosity monitor data from the first run indicated
that the luminosity had only been one eighth of what was expected from
the set flow rate. Inspection revealed a leak in the junction between
the tube that delivered the hydrogen gas to the cell assembly and
the cell itself. Most of the hydrogen was
leaking into the scattering chamber without ever
entering the target cell.  It was not possible to fix the leak in
situ, so the scattering chamber was removed.  A new cell was installed, with a
bayonet to bridge the gap between the two tube pieces. This
modification prevented any leak at this junction.

\subsection{OLYMPUS Run II}

The elimination of the leak increased the target thickness to the
design value, resulting in a
significant reduction of the DORIS beam lifetime compared to Run I. 
The short lifetime was overcome by operating in ``top-up'' mode. In this mode, the 
storage ring was refilled to the desired current every few minutes rather than allowing the beam current
to decay significantly. With 0.6~sccm flow, 
a typical beam lifetime was on the order of only ten minutes, but in
top-up mode, an average beam
current of more than 60~mA was sustained during data taking. At that flow, the pressure gauges 
at the central pumps reached a pressure of $2\times 10^{-6}$~Torr. The upstream pumps 
recorded pressures of approximately $1\times10^{-7}$ and $2\times 10^{-8}$~Torr. The downstream pumps 
recorded pressures of $5\times 10^{-7}$ and $3\times 10^{-8}$~Torr. Gauges just outside
the OLYMPUS beamline segment recorded pressures of $7\times 10^{-8}$~Torr and $2\times 10^{-8}$~Torr.
The average pressure in the DORIS ring was maintained at $3\times 10^{-9}$~Torr.

\begin{figure}[htb]
\centering
\includegraphics[width=\columnwidth]{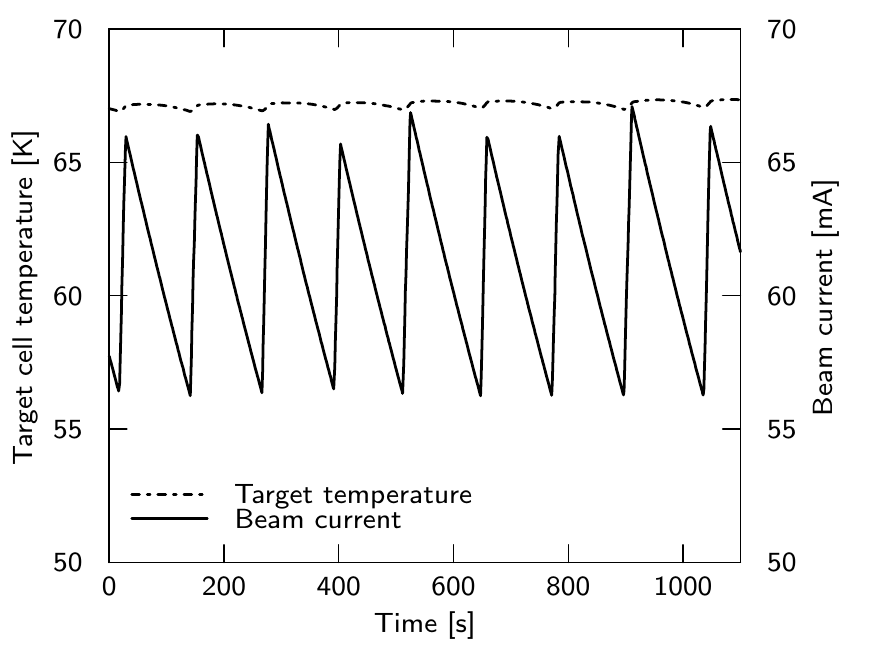}
\caption{During typical running, the target cell temperature was stably maintained in top-up mode, with only a small dependence on the beam current.}
\label{fig:temp}
\end{figure}

During the second run, a target temperature slightly below 70~K was maintained, as shown
in Fig.\ \ref{fig:temp}. Synchrotron operation was expected to produce more severe
wakefields because of the higher currents and fewer bunches,  but this turned out not 
to be the case. The higher temperature during OLYMPUS running was likely caused by the target enlarging the beam and increasing 
the wakefield heating. 

\begin{figure}[htb]
\centering
\includegraphics[width=\columnwidth]{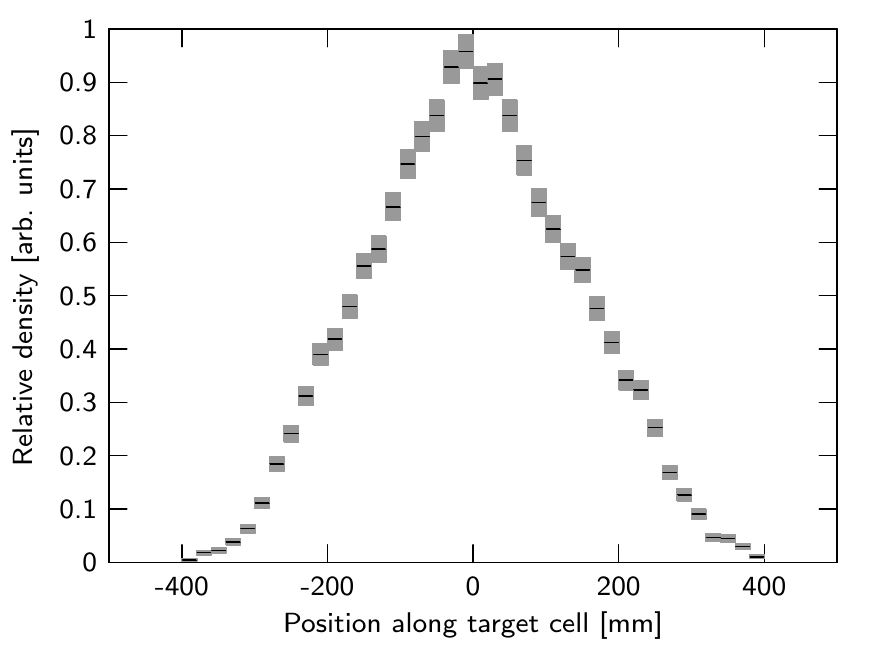}
\caption{The reconstructed target density distribution follows the expected
  triangular shape. An offset to negative positions and tails beyond
  the expected range of $\pm$300 mm are visible.}
\label{fig:zdis}
\end{figure}

In a preliminary analysis, it was possible to reconstruct the target
density distribution from a small subset of the full data set. The
result is shown in Fig.\ \ref{fig:zdis}. The distribution mainly follows
the expected triangular shape but is slightly shifted to negative positions,
i.e.\ upstream. The distribution is also slightly
wider than the cell. Further analysis will examine whether this is
real or an effect of the preliminary tracking and event selection.

\section{Conclusion}

The design and realization of the OLYMPUS target system has been presented. The OLYMPUS 
experiment required a unpolarized hydrogen gas target that 
could withstand a harsh beam environment. The target used an
internal gaseous design to avoid degrading the DORIS beam. 
The cell was cryogenically cooled to increase the target thickness
for a given flow. With significant wakefield suppression, the target
performed to specifications, and a temperature of 
under 70~K was sustained during operation. The analysis of data taken during
the OLYMPUS experiment is ongoing.

\section*{Acknowledgements}

We would like to thank Chris Vidal as well as the MIT-Bates technicians for helping 
to realize and install the OLYMPUS target. Additionally, we thank the members of 
the DORIS vacuum group for their assistance. We gratefully acknowledge the input 
from Rainer Wanzenberg and Uli K\"otz, whose correspondence regarding wakefields 
and beam halo were instrumental in the design of the target and collimator. We 
thank Uwe Schneekloth for coordinating the various efforts at DESY and for his 
helpful suggestions during the course of the target design and modification. We 
gratefully thank our other collaborators in the OLYMPUS experiment, who made the 
successful completion of the experiment possible and collected the data used to 
create Fig.\ \ref{fig:zdis}. We especially would like to express our gratitude to 
Alexander Winnebeck and J\"urgen Diefenbach who lent us invaluable assistance on 
numerous occasions. 

This work was supported by the Office of Nuclear Physics of the U.S.\ Department of Energy.





\bibliographystyle{model1-num-names}
\bibliography{references.bib}







\end{document}